\documentclass[showpacs,preprintnumbers,aps,prd,nofootinbib]{revtex4}
\usepackage[dvips]{graphicx}
\usepackage{bm,latexsym,amsmath,amssymb,amsfonts}

\def\az{\!\stackrel{\hbox{\tiny (0)}}{a}\!\!}
\def\da{{\delta a}}
\def\Hz
{\!\stackrel{\hbox{\tiny (0)}}{\cal H}\!{}}
\def\dH{{\delta {\cal H}}}
\def\phiz
{\!\stackrel{\hbox{\tiny (0)}}{\phi}\!{}}
\def\tildephiz
{\!\stackrel{\hbox{
$\sim\hspace{-11pt}{\lower3pt\hbox{\tiny(0)}}$}}\phi\!{}}
\def\uz
{\!\stackrel{\hbox{\tiny (0)}}{u}\!{}}
\def\Ez
{\!\stackrel{\hbox{\tiny (0)}}{\cal E}\!{}}

\begin{document}
%----------------%

\title{
Leading order corrections to the cosmological evolution of 
tensor perturbations in braneworld
}

\author{Tsutomu Kobayashi}
\email{tsutomu@tap.scphys.kyoto-u.ac.jp}
\author{Takahiro Tanaka} 
\email{tama@scphys.kyoto-u.ac.jp}

\affiliation{
Department of Physics, Kyoto University, Kyoto 606-8502, Japan 
}

\begin{abstract}
Tensor type perturbations in the expanding 
brane world of the Randall Sundrum type are investigated. 
We consider a model composed of slow-roll inflation phase 
and the succeeding radiation phase. The effect of the 
presence of an extra dimension through the transition 
to the radiation phase is 
studied, giving an analytic formula for leading 
order corrections. 
\end{abstract}


\preprint{KUNS-1933}

\maketitle

%-----------------------------------------%
\section{Introduction}
%-----------------------------------------%

In recent years there has been much interest in braneworld
scenarios~\cite{Rubakov:2001kp, Langlois:2002bb, Maartens:2003tw},
where our four dimensional universe
could be a brane embedded in a higher dimensional bulk spacetime.
The basic ingredient of braneworld scenarios is that
while ordinary matter fields are confined to the brane,
only gravitational interactions can propagate in the bulk.
The second Randall-Sundrum model (RS II)~\cite{Randall:1999vf}
with one brane in an anti de Sitter (AdS) bulk
is particularly interesting
in the point that, despite an infinite extra dimension,
four dimensional general relativity (4D GR) can be recovered
at low energies/long distances on the brane.
Then, what is leading order corrections to
the conventional gravitational theory?
It was shown that
the Newtonian potential in the RS II braneworld,
including the correction due to the bulk gravitational effects
with a precise numerical factor~\cite{Garriga:1999yh,Gidding}, is given by
\begin{eqnarray*}
V(r)\simeq -\frac{Gmm'}{r}\left(1+\frac{2\ell^2}{3r^2}\right).
\end{eqnarray*}
Here $\ell$ is the curvature length of the AdS space
and is experimentally constrained to be $\ell \lesssim 0.1$ mm.
It is natural to ask next what are leading order corrections
to cosmological perturbations.

Great efforts have been paid for the problem of
calculating cosmological perturbations in braneworld scenarios
(see, for example,
Refs.~\cite{Maartens:1999hf, Mukohyama:2000ui, Kodama:2000fa,
Langlois:2000ia, Langlois:2000ph, Koyama:2000cc, Koyama:2001ct, Koyama:2003be,
Langlois:2000ns, Gorbunov:2001ge, Kobayashi:2003cn, Hiramatsu:2003iz, %GW
Easther:2003re, Battye:2003ks, Ichiki:2003hf, Ichiki:2004sx, %GW
Koyama:2004cf, Koyama:2003yz, Koyama:2003sb, Kobayashi:2003cb, Binetruy:2004dw}).
In order to correctly evaluate perturbations on the brane,
we need to solve bulk perturbations, which reduces to a problem
of solving partial differential equations with appropriate 
boundary conditions.
Hence it is not as easy as in the standard four dimensional 
cosmology. 
A pure de Sitter braneworld~\cite{Langlois:2000ns}, and its variations
such as ``junction'' models~\cite{Gorbunov:2001ge, Kobayashi:2003cn}
and a special dilatonic braneworld~\cite{Koyama:2003yz, Koyama:2003sb, Kobayashi:2003cb}
are the rare examples where the bulk perturbations are analytically solved.
In more generic cases,
one has to resort to a numerical
calculation~\cite{Hiramatsu:2003iz, Ichiki:2003hf, Ichiki:2004sx},
or some approximate
methods~\cite{Koyama:2003be, Easther:2003re, Battye:2003ks, Koyama:2004cf}.
There is another difficulty concerned with a physical, fundamental aspect
of the problem; we do not know how to specify appropriate initial conditions
for perturbations with bulk degrees of freedom.

In this paper,
we investigate cosmological tensor perturbations in
the RS II braneworld, and try to evaluate leading order corrections
to the 4D GR result analytically.
For this purpose, 
we make use of the reduction scheme to a four dimensional 
effective equation which
iteratively takes into account the effects of the bulk gravitational
fields~\cite{Tanaka:2004ig}. 
As a first step, we concentrate on perturbations 
which is initially at a super-horizon scale.

The paper is organized as follows.
In Sec.~II we define the setup of the problem discussed 
in the present paper more precisely. 
In Sec.~III we briefly 
review the derivation of the effective equation of motion
for tensor perturbations. 
In Sec.~IV, using the long wavelength approximation
we develop a method
for evaluating leading order corrections,  
and evaluate the corrections
considering slow-roll inflation and
the radiation stage which follows after inflation. 
The shortcoming of the long wavelength approximation is 
cured by introducing another method 
suitable for analyzing the radiation phase in Sec. IV.
Section V is devoted to summary and discussion.

\section{Setup of the problem and background cosmological model}

In this paper we investigate tensor perturbations 
in a Friedmann braneworld model with an warped (infinite) extra 
dimension~\cite{Randall:1999vf, Binetruy:1999ut, Binetruy:1999hy,
Mukohyama:1999qx, Ida:1999ui, Kraus:1999it}. 
To determine the spectrum of 
tensor perturbations observed at late time, we need to 
solve the perturbation equations with appropriate 
initial conditions. However, it is not an easy task to specify 
what initial conditions we should use in the context of 
braneworld cosmology. At an earlier stage of the evolution 
of the universe the physical wavelength of 
perturbation modes is much shorter than the bulk curvature 
scale. Then, the correction 
to the four dimensional standard evolution will not remain small.  
Therefore, in this regime it is not appropriate to discuss 
the dynamics of perturbation modes 
under the assumption that
a modification to the four dimensional 
standard one is small. 
Hence our strategy is to isolate the issue of initial conditions 
from the rest. 
In this paper we discuss a relatively easy part, {\it i.e.}, late time 
evolution after the effect of the fifth dimension becomes perturbative.

Our discussion will be restricted to 
the perturbation modes whose wavelength 
is initially much longer than the Hubble horizon scale. 
We analytically derive a formula for small corrections to 
the solution for the tensor perturbation equation 
due to the presence of the fifth dimension. 
We mainly focus on the behavior of a growing solution,  
since a growing solution will be more important than a decaying one 
in general.  
Later we briefly mention the decaying solution in 
Sec.~\ref{sec:conclusion}. 
%We also mention the perspective about the issue of initial conditions 
%there. 

The background spacetime that we consider is composed of 
a five dimensional AdS bulk, whose metric is
given in Poincar\'{e} coordinates as
\begin{eqnarray*}
ds^2=\frac{\ell^2}{z^2} \left(
dz^2-dt^2+\delta_{ij}dx^idx^j \right),
\end{eqnarray*}
with a Friedmann brane
at $z=z(t)$.
Here $\ell$ is the curvature length of AdS space.
In the original RS II model~\cite{Randall:1999vf},
a brane is placed at a position of fixed $z$,
and Minkowski geometry is realized on the brane. 
The induced metric on $z=z(t)$ is
\begin{eqnarray}
ds^2&=&a^2(t)\left[-(1-\dot z^2)dt^2+\delta_{ij}dx^idx^j\right],
\label{induced_metric}\\
a(t)&=&\ell / z(t)\nonumber,
\end{eqnarray}
where the overdot denotes $\partial_t$.
From Eq.~(\ref{induced_metric}), we see that the conformal time on the brane
is given by $d\eta = \sqrt{1-\dot z^2} dt$.
The brane motion is related to the energy density of matter 
localized on the brane by the modified Friedmann
equation~\cite{Binetruy:1999ut, Binetruy:1999hy, Mukohyama:1999qx, Ida:1999ui, Kraus:1999it}
as
\begin{eqnarray}
H^2=\frac{8\pi G}{3}\left( \rho + \frac{\rho^2}{2\sigma} \right),
\label{FriedmannEq}
\end{eqnarray}
and
\begin{eqnarray}
H:={\dot a\over a}=\frac{-\ell^{-1}\dot z}{\sqrt{1-\dot z^2}},
\end{eqnarray}
where $\sigma =3/4\pi G \ell^2$ is a tension of the brane.
The quadratic term in $\rho$
in the Friedmann equation
modifies the standard cosmological expansion law.
At low energies ($\ell H \ll 1$),
the brane motion is non-relativistic ($\dot z \ll 1$),
and hence the conformal time on the brane $\eta$
almost coincides with the bulk conformal time $t$.

As a background Friedmann brane model, we consider 
slow-roll inflation at low energies 
followed by a radiation dominant phase. 
During the slow-roll inflation the wavelength of perturbations stays 
outside the Hubble scale, and it re-enters the Hubble horizon 
only during the radiation era.

We fix a model of slow-roll inflation by 
simply providing a function 
\begin{eqnarray}
\epsilon (\eta):=
1-\frac{\partial_{\eta}{\cal H}}{{\cal H}^2},  
\label{defepsilon}
\end{eqnarray}
where ${\cal H}:=\partial_\eta a/a= aH$. 
During slow-roll inflation, $\epsilon$ is supposed to be small.  
In the $\epsilon\to 0$ limit, we recover the de Sitter case
\begin{equation}
a= \frac{a_0\eta_0}{2\eta_0-\eta},\qquad
{\cal H}= \frac{1}{2\eta_0-\eta},\qquad(\eta<\eta_0), 
\label{deSitter}
\end{equation}
where we have chosen the normalization of the scale factor 
and the origin of the $\eta$-coordinate so that $a=a_0$ 
and ${\cal H}={1/\eta_0}$ at $\eta=\eta_0$. 

In the above 
alternatively we could specify the potential of 
the inflaton field localized on the brane.  
Then, for a given inflaton potential, 
the evolution of the background scale factor would differ from the 
conventional four dimensional one. 
However, as we will show below, the correction to the 
tensor perturbations coming from the slow-roll inflation phase 
is always small irrespective of the details of the evolution 
of the scale factor. 
Hence, we simply give time-dependence of the 
slow-roll parameter to specify a model. 

We assume that 
slow-roll inflation terminates at around $\eta=\eta_0$, 
and a radiation era follows.  
During the radiation era the energy density decreases 
like $\rho=\rho_0(a_0/a)^4$ as usual. 
Then, solving Eq.~(\ref{FriedmannEq}) at low energies, 
we find that the scale factor behaves like 
\begin{eqnarray}
a(\eta)=\az (\eta)+\delta a(\eta)+{\cal O}(\ell^4),
\qquad(\eta>\eta_0),
\label{scale_rad}
\end{eqnarray}
with 
\begin{eqnarray}
 \az (\eta) = a_0 {\eta\over\eta_0},
\qquad
 \da (\eta) = {a_0}\ell^2 H_0^2
    \left[\frac{1}{6}-\frac{1}{24}\left(\frac{\eta_0}{\eta}\right)^3
-\frac{1}{8}\left(\frac{\eta}{\eta_0} \right)   \right],
\label{scale_rad2}
\end{eqnarray}
and $H_0=(a_0\eta_0)^{-1}$.
Correspondingly, we have 
\begin{eqnarray}
{\cal H}(\eta) & = & 
  \Hz(\eta)+\dH(\eta)+{\cal O}(\ell^4),
\label{H_rad}
\end{eqnarray}
with 
\begin{eqnarray}
\Hz (\eta)= \frac{1}{\eta},\qquad 
\dH (\eta) 
  = {\ell^2\over 6a^2\eta_0^3}\left[
    \left({\eta_0\over \eta}\right)^3-1
            \right],
\qquad(\eta>\eta_0). 
\label{H_rad2}
\end{eqnarray}
It is convenient to extend the definition of 
the slow-roll parameter $\epsilon$ 
to the later epoch by 
\begin{eqnarray}
\epsilon (\eta):=
1-\frac{\partial_{\eta}{\Hz}}{\Hz^2}. 
\label{defepsilon2}
\end{eqnarray}

%-----------------------------------------%
\section{Tensor perturbations on a Friedmann brane}
%-----------------------------------------%

We use the method to reduce the five dimensional equation 
for tensor perturbations in a Friedmann 
braneworld
at low energies to a four dimensional effective equation of motion, 
derived in Ref.~\cite{Tanaka:2004ig}.
Here we summarize the derivation.

Tensor perturbations on a Friedmann
brane~\cite{Langlois:2000ns, Gorbunov:2001ge,
Kobayashi:2003cn, Hiramatsu:2003iz, Easther:2003re, Battye:2003ks, Ichiki:2003hf,
Ichiki:2004sx, Koyama:2004cf}
are given by
\begin{eqnarray}
ds^2=\frac{\ell^2}{z^2} \left[
dz^2-dt^2+(\delta_{ij}+h_{ij})dx^idx^j \right].
\end{eqnarray}
We expand the perturbations by
using $Y^k_{ij}({\bm x})$, a transverse traceless tensor harmonics with 
comoving wave number $k$, 
as $h_{ij}=\sum_{k}Y^k_{ij}({\bm x})\Phi_k(t,z)$. 
Then the equation of motion for the tensor perturbations in the bulk 
is given by
\begin{eqnarray}
\left( - \partial_z^2+\frac{3}{z} \partial_z+\partial_t^2+k^2 \right)
\Phi_k=0,
\end{eqnarray}
Hereafter we will discuss each Fourier mode separately, 
and we will abbreviate the subscript $k$.
The general solution of this equation is 
\begin{eqnarray}
\Phi
&=&\int d\omega \tilde \Psi(\omega) e^{-i\omega t}
2(pz)^2K_2(pz) \nonumber\\
&=&\int d\omega\tilde \Psi(\omega) e^{-i\omega t}
\left\{1-\frac{(pz)^2}{4}+\frac{(pz)^4}{16}
\left[\frac{3}{4}-\gamma-\ln\left(\frac{pz}{2}\right)\right]+\cdots
\right\},
\end{eqnarray}
where $p^2=-\omega^2+k^2$,
and $\gamma$ is Euler's constant. We have chosen
the branch cut of the modified Bessel function $K_2$ so that
there is no incoming wave from past null infinity in the bulk.
The coefficients $\tilde \Psi(\omega)$ are to be determined
by the boundary condition on the brane
$n^{\mu}\partial_{\mu}\Phi|_{z=z(t)}=0$, where $n^{\mu}$
is an unit normal to the brane, or, equivalently,
by the effective Einstein equations on the brane~\cite{Shiromizu:1999wj},
\begin{eqnarray}
^{(4)}G_{\mu\nu}=8\pi G\:T_{\mu\nu}+(8\pi G_5)^2 \pi_{\mu\nu}
-E_{\mu\nu}.
\end{eqnarray}
Here $\pi_{\mu\nu}$ is quadratic in the energy momentum tensor $T_{\mu\nu}$,
and the first two terms on the right hand side 
are totally represented by the variables which reside on the 
brane. 
A projected Weyl tensor $E_{\mu\nu}:=C_{\alpha\mu\beta\nu}n^{\alpha}n^{\beta}$
represents the effects of
the bulk gravitational fields, giving rise to corrections to
four dimensional Einstein gravity in a fairly nontrivial way.
In the present case, this can be written explicitly,
and the effective four dimensional equation reduces to
\begin{eqnarray}
\left( \partial_{\eta}^2+2{\cal H}\partial_{\eta}+k^2\right) \phi=-2E,\hspace{40mm}
\\
-2E=\left.\left[
(\ell H)^2(\partial_t^2+\partial_z^2)-2\ell H\sqrt{1+(\ell H)^2}\partial_t\partial_z
+\partial_z^2-\frac{1}{z}\partial_z \right]\Phi\right|_{z=z(t)},
\end{eqnarray}
where $\phi(t):=\Phi(t,z(t))$ is the perturbation evaluated on the brane.
At low energies ($\ell H \ll 1$), we can use the approximation like
\begin{eqnarray*}
\partial_t^2 \Phi \simeq \partial_{\eta}^2 \Phi,
~~~\partial_t\partial_z \Phi \simeq -\frac{\ell}{2a}\partial_{\eta}
(\partial_{\eta}^2+k^2)\Phi,
~~~\partial_z^2\Phi \simeq -\frac{1}{2}\int d\omega\tilde\Psi e^{-i\omega\eta}p^2
\simeq -\frac{1}{2}(\partial_{\eta}^2+k^2)\Phi,
\end{eqnarray*}
neglecting higher order corrections of ${\cal O}(\ell^4)$.
Then, using the lower order equation to eliminate the higher derivative terms
in $-2E$, we finally obtain our basic equation:
\begin{eqnarray}
\left( \partial_{\eta}^2+2\Hz \partial_{\eta}+k^2\right) \phi
\simeq \frac{\ell^2}{\az^2}\left(
 {\cal S}_0[\phi]+{\cal S}_1[\phi] + {\cal S}_2[\phi] 
\right),
\label{basic}
\end{eqnarray}
where
\begin{eqnarray}
{\cal S}_0[\phi]&=&-\frac{2\az^2}{\ell^2}\dH \phi', 
\label{Source3}
\\
{\cal S}_1[\phi]&=&
(3\Hz ^3-2\Hz \Hz')\phi' +k^2\Hz ^2
\phi,\label{Source1}
\\
{\cal S}_2[\phi]&=&-\frac{1}{2}\int d\omega
p^4\tilde \phi e^{-i\omega \eta}
\left[ \ln \left(\frac{p\ell}{2a}\right) +\gamma \right],
\label{Source2}
\end{eqnarray}
and the prime denotes $\partial_{\eta}$.
The first term arises due to the non-standard cosmological expansion 
included in ${\cal H}$, while the last two terms are 
the corrections from the bulk effects $E_{\mu\nu}$.
We can see that all terms 
are suppressed by $\ell^2$ (or $\ell^2\ln\ell$)
\footnote{Hereafter, we will refer to both terms 
suppressed by $\ell^2$ and by $\ell^2\ln\ell$ as ${\cal O}(\ell^2)$.}.
${\cal S}_2$ is essentially nonlocal because of 
the presence of the log term. 
A time-domain expression for the quantity $p^4 \phi$ 
which appears in ${\cal S}_2$ can be rewritten, 
by eliminating the higher derivative terms
in a similar way as before, as 
\begin{eqnarray}
\hat p^4 \phi \simeq -2\left[ \left(
\Hz''-6\Hz \Hz'+4\Hz ^3
\right) \partial_{\eta} +2k^2 \left(
\Hz ^2-\Hz'\right) \right]\phi,
\end{eqnarray}
where we have introduced a derivative operator
$\hat p^2 = \partial_{\eta}^2+k^2$.

%-----------------------------------------%
\section{New long wavelength approximation}
%-----------------------------------------%
Let us begin with the case that the wavelength of 
perturbations is much longer than the Hubble scale. 
The long wavelength approximation
not only allows us an easy treatment of perturbations,
but also turns out to be sufficient to derive 
all the major corrections except for that coming from 
${\cal S}_0$. 
We would like to stress that 
what we call the long wavelength approximation 
here is different from a low energy expansion scheme
or a gradient expansion scheme
in the literature~\cite{KannoSoda, Kanno:2002ia, Koyama:2002nw}.

%-----------------------------------------%
\subsection{General iteration scheme}
%-----------------------------------------%

The equation we are considering takes the form of  
\begin{eqnarray}
\left( \partial_{\eta}^2+2\Hz \partial_{\eta}+k^2\right) \phi
= \frac{\ell^2}{\az^2}{\cal S}[\phi]. 
\label{iterationeq}
\end{eqnarray}
This can be solved iteratively taking $\ell^2$ as a small parameter.
For this purpose, we write
\begin{eqnarray}
\phi(\eta)=
\phiz(\eta)e^{F(\eta)}=
\phiz(\eta)\exp\left[\int^{\eta}f(\eta') d\eta' \right].
\end{eqnarray}
A zeroth order solution $\phiz$
by definition satisfies 
the equation obtained by setting the right hand side 
of Eq.~(\ref{iterationeq}) to be zero. 
Then $f$ obeys
\begin{eqnarray}
\partial_{\eta}f+2\left({\partial_\eta\phiz\over \phiz}+\Hz \right)f
= \frac{\ell^2}{\az^2 \phiz}{\cal S}[\phiz],
\end{eqnarray}
which can be integrated immediately to give
\begin{eqnarray}
f(\eta) = \frac{\ell^2}{\az^2\phiz^{2}}
\int^{\eta}
d\eta' \phiz{\cal S}[\phiz].
\end{eqnarray}
Integrating this expression, we obtain the first order correction $F$.
In the present case, ${\cal S}$ consists of three parts, 
${\cal S}_0$, ${\cal S}_1$ and ${\cal S}_2$.
We hereafter denote the corrections $f$ and $F$ 
coming from ${\cal S}_i$ as $f_i$ and $F_i$, respectively.

\subsection{Long wavelength expansion of the zeroth order solution}

In the long wavelength approximation,
we can write the zeroth order solution explicitly as
\begin{eqnarray}
\phiz&\simeq&
A_k\left[1-k^2\int^{\eta}d\eta'I(\eta')\right],
\label{zeroth-O-sol}\\
\partial_{\eta}\phiz&\simeq&-A_kk^2I(\eta),
\end{eqnarray}
where $A_k$ is an amplitude of the perturbation 
which can depend on $k$, and
\begin{eqnarray}
I(\eta):={\az^{-2}(\eta)}\int^{\eta}_{-\infty} \az^2(\eta')d\eta'.
\end{eqnarray}
Here we simply keep the terms up to ${\cal O}(k^2)$. 

We have already introduced a parameter 
$\epsilon(\eta)$ in %Eq.~(\ref{defepsilon}) or 
Eq.~(\ref{defepsilon2}), 
which parametrizes the cosmological expansion. 
In the inflationary universe,
it reduces to a slow-roll parameter and is assumed to be small.
Slow-roll inflation assumes that 
time differentiation of $\epsilon$ is ${\cal O}(\epsilon^2)$. 
Hence, we treat 
$\epsilon$ as a constant during slow-roll inflation. 
The transition from inflation
to the radiation stage occurs at
around $\eta=\eta_0$,
and the behavior of $\epsilon$ there depends on
the details of the reheating process.
During the radiation stage, $\epsilon$ keeps again a constant value,
$\epsilon = 2$,
because the scale factor behaves like $\az \;\propto \eta$.
In general $\epsilon$ reduces to a constant 
for the scale factor $\az$ proportional to
a power of the conformal time.

Using $\epsilon$ and
the zeroth order solution in the long wavelength
approximation,
we can write ${\cal S}_1[\phiz]$ 
and $\hat p^4\phiz$
as
\begin{eqnarray}
{\cal S}_1[\phiz]&=&\Hz ^2\left[
(1+2\epsilon)\Hz \partial_{\eta}\phiz+k^2\phiz \right]
\nonumber\\
&\simeq&A_kk^2\Hz ^2\left[
1-\Hz I-2\epsilon\Hz I \right],
\label{S1slowrollpara}
\\
\hat p^4 \phiz &\simeq&
2\Hz ^2\left\{\left[ \epsilon'-2\epsilon(1+\epsilon)\Hz \right]
\partial_{\eta}\phiz-2\epsilon k^2\phiz \right\}
\nonumber\\
&\simeq&
-2 A_kk^2\Hz ^2\left\{
\epsilon' I-2\epsilon\left[(1+\epsilon)\Hz I-1\right]
\right\},
\label{p4phiep}
\end{eqnarray}
which are generic expressions
not relying on any specific time dependence of the scale factor.

Equation~(\ref{p4phiep}) can be further rewritten into a more suggestive form.
First,
integrating the identity $\partial_{\eta}(\az^2/\Hz )=(1+\epsilon)\az^2$,
it can be shown that
\begin{eqnarray*}
1-\Hz I = \frac{\Hz  }{\az^2}\int^{\eta}_{-\infty}\epsilon \az^2 d\eta'.
\end{eqnarray*}
Then, integration by parts leads
\begin{eqnarray}
(1+\epsilon)\Hz I-1=
\frac{\Hz  }{\az^2}\int^{\eta}_{-\infty}\epsilon' \az^2 I d\eta',
\label{int_by_parts}
\end{eqnarray}
and thus we have
\begin{eqnarray}
\hat p^4 \phiz \simeq
-2 A_kk^2\Hz ^2\left(
\epsilon' I-2\epsilon\frac{\Hz  }{\az^2}\int^{\eta}_{-\infty}\epsilon' \az^2 I d\eta'
\right).
\label{ep_dash_p4}
\end{eqnarray}
Since $\epsilon'$ is of second order
in the slow-roll parameters during slow-roll inflation, 
we can safely neglect this term until the end of inflation.
Moreover, 
for sufficiently smooth $\epsilon$
this term is expected to be small, and
$\epsilon'=0$ for the scale factor proportional to
a power of the conformal time.
Only at the time when the equation of state of the universe
changes abruptly, {\em i.e.}, at a sudden transition from inflation to
the radiation stage,
$\epsilon'$ possibly becomes significantly large like a delta function.
After such a violent transition,
the integral in Eq.~(\ref{ep_dash_p4})
leaves a constant value and thus this second term decreases
in proportion to the damping factor $\Hz /\az^2$.

%-----------------------------------------%
\subsection{Slow-roll inflation}
%-----------------------------------------%

Now let us consider slow-roll inflation on the brane.
In this case ${\cal S}_0$ vanishes, since 
$\dH=0$ by construction. 
In the previous section we have also shown that
$\hat p^4 \phiz$ vanishes at first order in the slow-roll parameters
during inflation.
Consequently, the only relevant term
in the present situation is ${\cal S}_1$.

Equation~(\ref{S1slowrollpara}) together with Eq.~(\ref{int_by_parts}) tells us that
\begin{eqnarray}
{\cal S}_1\simeq -A_kk^2\Hz ^2\epsilon,
\end{eqnarray}
and thus
\begin{eqnarray}
f(\eta)
&\simeq&
-\frac{k^2\ell^2}{\az^2}\int^{\eta}d\eta'\epsilon\Hz ^2.
\end{eqnarray}
Using the relations $\Hz \simeq \az/a_0\eta_0$ and 
$d\eta=a_0\eta_0\,d\!\az/\az^2$, 
and neglecting the time variation of the slow-roll parameter,
we obtain
\begin{eqnarray}
f(\eta)\simeq 
\frac{k^2\ell^2}{\eta_0 \!\az^2}\epsilon\left(
 \frac{a_i}{a_0}-\frac{a}{a_0}\right),
\end{eqnarray}
where $a_i=a(\eta_i)$ and 
$\eta_i$ is the lower boundary of integration. 
Integrating this expression, we have
\begin{eqnarray}
F(\eta)\simeq -
\frac{k^2\ell^2}{a_i^2}\epsilon\left[
\frac{1}{3}\left(\frac{a_i}{\az}\right)^3-\frac{1}{2}
 \left(\frac{a_i}{\az}\right)^2
\right].
\label{Feta}
\end{eqnarray}
We chose the integration constant so that
$F$ vanishes in the limit of $a(\eta)\to \infty$.  
This means that we have renormalized $A_k$ 
so that it represents the amplitude of fluctuations that we 
see at late times if slow-roll inflation lasted forever.

Although our present long wavelength approximation 
is not valid when the wavelength $a_i k^{-1}$ 
becomes comparable or shorter 
than the Hubble scale $H$,  
we can still arrange $a_i$ to be sufficiently small 
so that we can neglect 
the first term in Eq.~(\ref{Feta}).  
For such $a_i$, we have an expression independent of $a_i$, 
\begin{eqnarray}
F(\eta) \approx \frac{k^2\ell^2}{2 \az^2}\epsilon,  
\end{eqnarray}
and we do not need to care about the choice of the ``initial time'' $\eta_i$.
We see that the correction arising during inflation
is very tiny, suppressed by the slow-roll parameter 
$\epsilon$ in addition to the factor $k^2\ell^2/\az^2$.
For pure de Sitter inflation there is no correction, as is expected.

%-----------------------------------------%
\subsection{Transition to the radiation stage}
%-----------------------------------------%

\begin{figure}[t]
  \begin{center}
    \includegraphics[keepaspectratio=true,height=55mm]{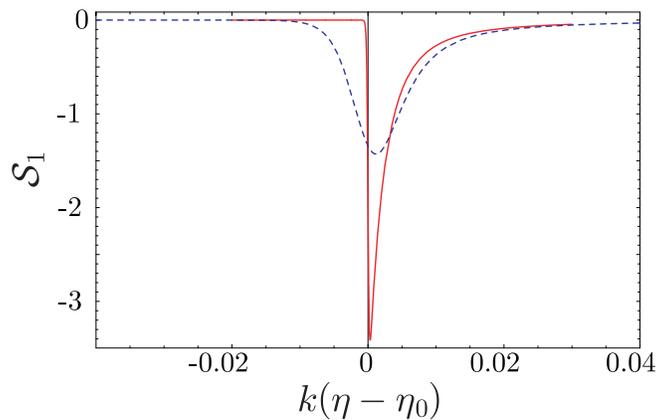}
  \end{center}
  \caption{Behavior of ${\cal S}_1$ around the transition time
  with the vertical axis in an arbitrary unit.
  Red solid line shows the case of a sharp transition
  with $s=0.02\eta_0$, 
% where
%  $\Hz _0$ is the comoving Hubble parameter at $\eta=\eta_0$,
  while blue dashed line represents the case of a smooth transition
  with $s=0.5\eta_0$.
  The wavelength of the mode is chosen to be $k\eta_0=0.01$.
%$\Hz (\eta_0)=1/\eta_0$. Thus, it is required
%that $k\eta_0 \ll 1$ for the long wavelength approximation to be valid,
%and indeed we have set $k\eta_0=0.01$.
  }
  \label{fig:S1.eps}
\end{figure}

\begin{figure}[t]
  \begin{center}
    \includegraphics[keepaspectratio=true,height=55mm]{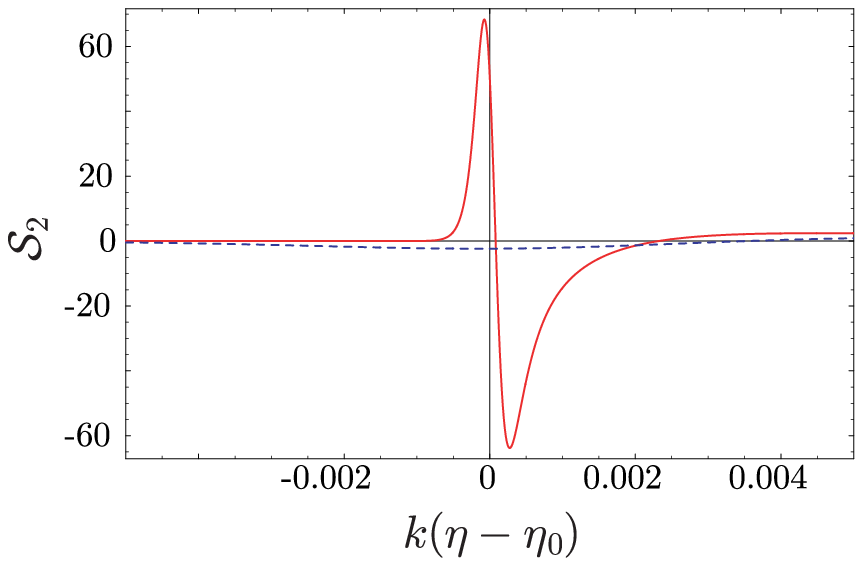}
  \end{center}
  \caption{Behavior of ${\cal S}_2$ around the transition time
  with the vertical axis in an arbitrary unit.
  Red solid line shows the case of a sharp transition
  with $s=0.02\eta_0$,
  while blue dashed line represents the case of a smooth transition
  with $s=0.5\eta_0$.
  The choice of the parameter is the same as in Fig.~\ref{fig:S1.eps},
  $k\eta_0=0.01$.}
  \label{fig:S2.eps}
\end{figure}

In this section, we investigate the effects of the transition from
the inflation stage to the radiation stage.
First, just for the illustrative purpose,
we plot the behavior of ${\cal S}_1$ and ${\cal S}_2$
for the modes well outside the horizon
in the neighborhood of the transition time $\eta_0$
in Fig.~\ref{fig:S1.eps} and Fig~\ref{fig:S2.eps}. 
These plots are for $\epsilon$ given by
\begin{eqnarray}
\epsilon(\eta)= \tanh[(\eta-\eta_0)/s]+1,
\end{eqnarray}
where the parameter $s$ controls the smoothness of the transition.
From the figures it can be seen that
the corrections from $E_{\mu\nu}$, 
${\cal S}_1$ and ${\cal S}_2$,
become significant only around the transition time.


From now on we shall consider the limiting case where
$\epsilon$ is given by $\epsilon(\eta)=2\theta(\eta-\eta_0)$.
Here we neglected the tiny effect of the non vanishing 
slow-roll parameter during inflation. 
For this instantaneous transition model,
the scale factor and the comoving Hubble parameter for 
$\eta<\eta_0$ are given by Eq.~(\ref{deSitter}), 
while for $\eta>\eta_0$ they are given by 
$\az$ and $\Hz$ in Eqs.~(\ref{scale_rad2}) and (\ref{H_rad2}).

Assuming such a step function-like transition,
we can evaluate the corrections from ${\cal S}_0$ 
at $\eta > \eta_0$ as 
\begin{eqnarray}
&&f_0(\eta) \simeq
-\frac{k^2\ell^2}{9\az^2}\int^{\eta}_{\eta_0}
\frac{\eta'd\eta'}{\eta^3_0}
   \left(1-\frac{\eta_0^3}{\eta^3}\right)
   \left(1+2\frac{\eta_0^3}{\eta^3}\right)
,\\
&&F_0(\eta) \simeq -\frac{k^2\ell^2}{90 a^2_0}
\left[5\frac{\eta}{\eta_0}-9+5\left(\frac{\eta_0}{\eta}\right)^2
-\left(\frac{\eta_0}{\eta}\right)^5
     \right].
\label{F11_an}
\end{eqnarray}
Here the integration constant is chosen so that
the correction vanishes before the transition.
From this expression, 
we find that $F_0$ diverges for a large $\eta$, 
and hence the correction to the final amplitude of fluctuations 
from ${\cal S}_0$ looks infinitely large. 
However, this is an artifact of the long wavelength approximation.
At a late time during the radiation era the zeroth order solution 
becomes oscillatory. 
This oscillation suppresses the late time contribution.  
However, in the long wavelength approximation, this
oscillation is not taken into account. 
In the succeeding section, we will develop 
another method which takes into account this oscillation. 
There we will find a finite correction. 

As for ${\cal S}_1$, we have 
\begin{eqnarray}
&&f_1(\eta) \simeq
-\frac{2}{3}\frac{k^2\ell^2}{\az^2}\int^{\eta}_{\eta_0} 
\frac{d\eta'}{\eta'{}^2}\left(1+5\frac{\eta_0^3}{\eta'{}^3}\right)
,\\
&&F_1(\eta) \simeq -\frac{2}{3}\frac{k^2\ell^2}{a^2_0}
\left[
\frac{3}{2}-\frac{9}{4}\left(\frac{\eta_0}{\eta}\right)
+\frac{1}{2}\left(\frac{\eta_0}{\eta}\right)^2
     +\frac{1}{4}\left(\frac{\eta_0}{\eta}\right)^5
\right].
\label{F11_an}
\end{eqnarray}
Long time after the transition 
(but of course before horizon re-entry),
the correction becomes time independent,
\begin{eqnarray}
\lim_{\eta\to\infty }F_1(\eta) 
=
 -\frac{k^2\ell^2}{a^2_0}.
\label{corr_from_local}
\end{eqnarray}

\begin{figure}[b]
  \begin{center}
    \includegraphics[keepaspectratio=true,height=50mm]{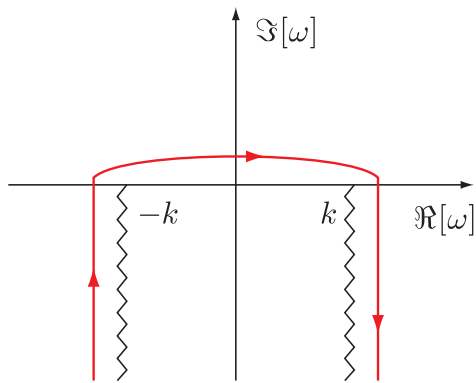}
  \end{center}
  \caption{Branch cuts and the contour of the integration
  in Eq.~(\ref{omega-int})
  for $T>0$.
  For $T<0$ we close the contour on the upper half plane.}
  \label{fig:cont1.eps}
\end{figure}

The next step is to evaluate the correction coming from ${\cal S}_2$.
This can be done as follows.
First, for the instantaneous transition
$\hat p^4\phiz$ becomes
\begin{eqnarray}
\hat p^4\phiz \simeq -4A_k k^2 \left[
\frac{1}{\eta_0}\delta(\eta-\eta_0)
-4\theta(\eta-\eta_0)\frac{\eta_0^3}{\eta^5}\right]
=-4A_k k^2\partial_{\eta}\left[
\theta(\eta-\eta_0)\frac{\eta_0^3}{\eta^4}\right].
\end{eqnarray}
Then, integrating by parts,
the Fourier transform of $\hat p^4 \phiz$ is obtained as
\begin{eqnarray}
p^4\tildephiz &=&
\frac{2A_kk^2}{\pi}i\omega \int d\eta'
\theta(\eta-\eta_0)\frac{\eta_0^3}{{\eta'}^4}e^{i\omega\eta'}
\nonumber\\
&=&\frac{2A_kk^2}{\pi}i\omega e^{i\omega\eta_0} q(\omega\eta_0),
\end{eqnarray}
with
\begin{eqnarray}
q(\omega\eta_0):=\int^{\infty}_{1}dy\frac{e^{i\omega\eta_0(y-1)}}{y^4}.
\end{eqnarray}
We separate ${\cal S}_2$ into two parts,
\begin{eqnarray*}
{\cal S}_2 = -\frac{1}{4}\int d\omega p^4 \tildephiz e^{-i\omega\eta}
\ln\left(\frac{p^2}{k^2}\right)-\frac{1}{2}\left[
\gamma+\ln\left(\frac{k\ell}{2a_0}\right)
+\ln\left(\frac{\eta_0}{\eta}\right)\right]\hat p^4\phiz,
\end{eqnarray*}
and, by substituting the expressions obtained above, we have
\begin{eqnarray}
{\cal S}_2 = \frac{A_kk^2}{2\pi}\int d\omega~ q(\omega\eta_0)
\partial_{\eta} \left[e^{-i\omega T}
\ln\left(\frac{p^2}{k^2}\right)\right]
+2A_kk^2 \left[
\gamma+\ln\left(\frac{k\ell}{2a_0}\right)
+\ln\left(\frac{\eta_0}{\eta}\right)\right]
\partial_{\eta}\left[\theta(T)\frac{\eta_0^3}{\eta^4}\right],
\end{eqnarray}
where $T:=\eta-\eta_0$.
Then, it can be integrated to give
\begin{eqnarray}
f_2(\eta) &=& \frac{k^2\ell^2}{\az^2}\biggl\{
\frac{1}{2\pi}\int d\omega~ q(\omega\eta_0)
\left[e^{-i\omega T}
\ln\left(\frac{p^2}{k^2}\right)\right]
\nonumber\\
&&~~+2\left[
\gamma+\ln\left(\frac{k\ell}{2a_0}\right)
+\ln\left(\frac{\eta_0}{\eta}\right)\right]
\theta(T)\frac{\eta_0^3}{\eta^4}+
\int^{\eta}_{-\infty}d\eta'
\theta(\eta'-\eta_0)\frac{2}{\eta'}\frac{\eta_0^3}{{\eta'}^4}
\biggr\}.
\label{omega-int}
\end{eqnarray}
The brunch cuts of $\log$ function and
the path of $\omega$ integration in the first term
should be chosen so that the retarded boundary condition is ensured,
which are presented in Fig.~\ref{fig:cont1.eps}.
The integration is dominated by $\omega^2\approx k^2$,
where we may approximate
$q(\omega\eta_0) \approx q(k\eta_0) \approx 1/3$
because $k\eta_0$ is assumed to be small.
Then, the first term can be written by using the formula presented in Appendix,
and can be integrated to give a part of $F_2(\eta)$ as
\begin{eqnarray*}
&&-\frac{2k^2\ell^2}{3a_0^2}\int^{\eta}_{-\infty}d\eta'\frac{\eta_0^2}{{\eta}'^2}\left\{
\theta(\eta'-\eta_0)\frac{\cos[k(\eta'-\eta_0)]-1}{\eta'-\eta_0}
+\partial_{\eta'}[\theta(\eta'-\eta_0)\ln (k(\eta'-\eta_0))]
+\gamma\delta(\eta'-\eta_0)\right\}
\\
&&\simeq
-\frac{2k^2\ell^2}{3a_0^2}\left\{
\frac{\eta_0^2}{\eta^2}\ln (k(\eta-\eta_0))
+\ln (k\eta_0)\left(1-\frac{\eta_0^2}{\eta^2}\right)
-1+\frac{\eta_0}{\eta}+\ln\left(\frac{\eta-\eta_0}{\eta}\right)
-\frac{\ln(\eta/\eta_0-1)}{(\eta/\eta_0)^2}+\gamma
\right\}
\\
&&\mathop{\longrightarrow}_{\eta\to\infty}
-\frac{2k^2\ell^2}{3a_0^2}\left[
\gamma-1+\ln (k\eta_0)\right],
\end{eqnarray*}
where we have neglected the term proportional to $\cos[k(\eta'-\eta_0)]-1$
in the integrand
because this term brings the higher order contributions in $k$.
The other part of $F_2(\eta)$ is obtained
by integrating the remaining parts in Eq.~(\ref{omega-int}) as
\begin{eqnarray*}
&&\frac{2k^2\ell^2}{a_0^2}
\left\{\left[\gamma+\ln\left(\frac{k\ell}{2a_0}\right)\right]
\frac{1}{5}\left(1-\frac{\eta_0^5}{\eta^5}\right)
+\frac{1}{25}\left[
\frac{\eta_0^5}{\eta^5}-1+5\frac{\ln(\eta/\eta_0)}{(\eta/\eta_0)^5}
\right]
+\frac{1}{4}\left(1-\frac{\eta_0}{\eta}\right)
-\frac{1}{20}\left(1-\frac{\eta_0^5}{\eta^5}\right)
\right\}
\\
&&\mathop{\longrightarrow}_{\eta\to\infty}
\frac{2k^2\ell^2}{a_0^2}\left\{\frac{1}{5}
\left[\gamma+\ln\left(\frac{k\ell}{2a_0}\right)\right]+\frac{4}{25}
\right\}.
\end{eqnarray*}
Combining the above two, we finally obtain the correction
from ${\cal S}_2$,
\begin{eqnarray}
\lim_{\eta\to \infty}F_2(\eta)=
\frac{2k^2\ell^2}{a_0^2}\left\{
\frac{37}{75}-\frac{2}{15}\left[\gamma+\ln\left(
\frac{k}{a_0H_0}
\right)\right]
+\frac{1}{5}\ln\left(\frac{\ell H_0}{2}\right)
\right\}.
\label{F2fin}
\end{eqnarray}

%-----------------------------------------%
\section{Corrections in the radiation stage}
%-----------------------------------------%

So far we have worked in the long wavelength approximation,
and found $k$-dependent small corrections 
to the spectrum of tensor perturbations due to 
${\cal S}_1$ and ${\cal S}_2$, which correspond to 
$E_{\mu\nu}$ term.  
It might be possible, however, that
further corrections arise after the mode re-enters the horizon.
We will show that such corrections are highly suppressed.
A basic observation supporting this conclusion is that 
the contributions from $E_{\mu\nu}$ term in the right hand side of Eq.~(\ref{basic})
become significantly large only at the transition time.
It decreases in powers of $a(\eta)$ after the transition,
and will be negligible when the long wavelength approximation brakes down.

In the long wavelength approximation% developed in the preceding section
, we could not obtain a meaningful estimate 
for the leading order correction 
caused by unconventional cosmic expansion, 
${\cal S}_0$. Here 
we will give a more rigorous treatment, and  
will resolve this drawback of the 
approach taken in the preceding section. 

In the radiation stage,
\begin{equation}
u(\eta):=
        a(\eta)\phi(\eta), 
\end{equation}
is a convenient variable. 
In terms of this new variable, we can rewrite the equation of motion~(\ref{basic}) as
\begin{eqnarray}
u''+k^2u
=
\frac{\ell^2}{a} \left(\bar{\cal S}_0+ {\cal S}_1+{\cal S}_2\right),
\label{ham_osc}
\end{eqnarray}
with
\begin{equation}
 \bar{\cal S}_0=\frac{a''}{\ell^2}u. 
\end{equation}
If we neglect the deviation of the expansion law from the standard one,
we have $a\propto \eta$ in the radiation stage and so $a''=0$.
This means that the first term on the right hand side
gives a contribution of ${\cal O}(\ell^2)$ 
from the modification of the expansion law 
as ${\cal S}_0$ in the preceding section.
Thus, all the terms collected on the right hand side are of 
$O(\ell^2)$, 
and at zeroth order $u$ behaves just like a harmonic oscillator.
The zeroth order solution is expressed by a linear
combination of the standard modes:
\begin{eqnarray}
\uz(\eta)=A_k\left[
 \alpha \frac{e^{-ik\eta}}{\sqrt{2k}}
+\beta \frac{e^{ik\eta}}{\sqrt{2k}}\right],
\label{linear_combination}
\end{eqnarray}
where the coefficients $\alpha$ and $\beta$ are
determined by matching the solutions at the onset of the radiation
stage.

We define ``energy'' of the harmonic oscillator as
\begin{eqnarray}
{\cal E} := \frac{1}{2}\left| u'\right|^2+\frac{k^2}{2}|u|^2.
\end{eqnarray}
Then, this energy is conserved at the zeroth order, i.e.,
\begin{eqnarray*}
{\Ez}=\frac{k}{2}(|\alpha|^2+|\beta|^2)
   |A_k|^2=~\mbox{const.}
\end{eqnarray*}
Taking into account the corrections of ${\cal O}(\ell^2)$,
the total variation of the energy can be estimated as
\begin{eqnarray}
\Delta{\cal E}=\sum_{j=0}^2\Delta {\cal E}_j
=\sum_{j=0}^2\int_{\eta_0}^\infty {\cal E}'_j d\eta, 
\end{eqnarray}
with %$\Delta{\cal E}_j=\int_{\eta_0}^\infty {\cal E}'_j d\eta$, and
\begin{eqnarray}
{\cal E}'_{0}
 & =  &\frac{\ell^2}{2a}(u')^*\bar{\cal S}_0+~\mbox{c.c.}, \cr
&&
\cr 
{\cal E}'_{j}
 & = & \frac{\ell^2}{2a}(u')^*{\cal S}_j+~\mbox{c.c.},
\qquad \mbox{for}~j=1,2. 
\end{eqnarray}
The amplitude of the oscillator $|A_k| e^{\Re[F]}$ 
is proportional to ${\cal E}^{1/2}\az/a$. 
The factor $\az/a$ here is to be attributed to unconventional 
cosmic expansion. 
Hence we have 
\begin{eqnarray}
\lim_{\eta\to\infty}
  \left\{\Re [F_0(\eta)] +{\da(\eta)\over\az(\eta)}\right\}
  & = &  \frac{\Delta {\cal E}_0}{2{\Ez}}, 
\nonumber \\
\lim_{\eta\to\infty}
  \Re [F_j(\eta)] 
  & = &  \frac{\Delta {\cal E}_j}{2{\Ez}},
\qquad \mbox{for}~j=1,2. 
\label{Fformula}
\end{eqnarray}

\subsection{Corrections due to the unconventional expansion law}

We start with computing the correction due to 
the unconventional cosmic expansion, $F_0$, 
which we could not determine in the long wavelength approximation. 
To do so, we first need to determine the coefficients
in Eq.~(\ref{linear_combination}) by requiring that
$u^{(0)}$ and its derivative are continuous across
the transition from the de Sitter stage to the radiation stage.
We assume for a moment that the perturbation in the initial phase
is given by the de Sitter growing mode,
\begin{eqnarray}
v(\eta)\simeq {A_k a_0\eta_0\over 2\eta_0 -\eta}
  \left[{1}
   +{k^2(2\eta_0-\eta)^2\over 2}\right].
\label{dsmodefunc}
\end{eqnarray}
Solving
\begin{eqnarray*}
v(\eta_0)&=&\uz (\eta_0),\\
\partial_{\eta} v(\eta_0)&=&\partial_{\eta}\uz (\eta_0),
\end{eqnarray*}
we obtain the coefficients as
\begin{eqnarray}
\alpha\simeq{i a_0\over \eta_0\sqrt{2k}}
\left[1-\frac{2i}{3}(k\eta_0)^3 \right],
~~~\beta=\alpha^*.
\label{bog_c}
\end{eqnarray}
The energy of the harmonic oscillator at zeroth order
is approximately given by
\begin{eqnarray}
\Ez \simeq {|A_k|^2a_0^2\over 2\eta_0^2}. 
\end{eqnarray}

Next we integrate
\begin{eqnarray}
{\cal E}'_0&=&\frac{i}{4}\frac{a''}{a}(|\alpha|^2-|\beta|^2-\alpha\beta^*e^{-2ik\eta}+
\alpha^*\beta e^{2ik\eta})|A_k|^2+~\mbox{c.c.}
\nonumber\\
&=&\frac{a''}{a}\Im[\alpha\beta^*e^{-2ik\eta}]|A_k|^2
\label{E0}
\end{eqnarray}
The scale factor correct up to ${\cal O}(\ell^2)$ is given 
in Eq.~(\ref{scale_rad}). 
Using this expression,  we have 
\begin{eqnarray}
\frac{a''}{a}\simeq 
-\frac{\ell^2H_0^2}{2}\frac{\eta_0^4}{\eta^6}.  
\end{eqnarray}
Then, integrating Eq.~(\ref{E0}), we have
\begin{eqnarray}
\frac{\Delta {\cal E}_0}{2\Ez} &\simeq
&-{\ell^2 H_0^2\eta_0^2\over 2a_0^2} \Im \left[\alpha\beta^* 
\int^{\infty}_{\eta_0}d\eta
\frac{\eta_0^4}{\eta^6}e^{-2ik\eta} \right]
\cr
&\simeq &
{\ell^2 H_0^2\over 4k\eta_0} \Im \left[
 \left(1-{2i\over 3}(k\eta_0)^3\right)^2
\left[
 \frac{1}{5}-\frac{i}{2}k\eta_0-\frac{2}{3}(k\eta_0)^2
 +\frac{2i}{3}(k\eta_0)^3+{\cal O}\left((k\eta_0)^4\right)
\right]\right]
\cr
&\simeq &
-\frac{1}{8}\ell^2H_0^2+\frac{1}{10}\frac{k^2\ell^2}{a_0^2}.
\label{mod_ex_energy}
\end{eqnarray}
The first term in the last line is not suppressed for 
the $k\to 0$ limit. This is  
because this term arises due to the asymptotic behavior of $\da(\eta)$.
In fact one can see from Eq.~(\ref{Fformula}) that
the expression for 
the physical amplitude of the perturbations $e^{F_0}$ 
does not have this 
term, and finally we have 
\begin{equation}
\lim_{\eta\to\infty}\Re[F_0(\eta)]= 
\frac{1}{10}\frac{k^2\ell^2}{a_0^2}.
\label{F0fin}
\end{equation}

\subsection{Suppression at sub-horizon scales}

Here we shall show that corrections from the $E_{\mu\nu}$ term
are suppressed at sub-horizon scales and hence
the result obtained in the preceding section 
by using the long wavelength approximation
suffices for our purpose.

Substituting $\phiz=\uz/\az\;$ into ${\cal S}_1[\phiz]$,
energy gain or loss due to this term is obtained as
\begin{eqnarray}
{\cal E}'_1=\frac{\ell^2}{2\az^2}\left[
\frac{5}{\eta^3} | \uz' |^2+
\left(-\frac{5}{\eta^4}+\frac{k^2}{\eta^2}\right)(\uz')^*\uz
\right] + \mbox{c.c.}.
\end{eqnarray}
Note that $| \uz' |^2$ and $(\uz')^*\uz$
consist of just constant parts and
oscillating parts with constant amplitudes.
Therefore, it is manifest that when we integrate ${\cal E}'_1$ 
the dominant contribution
comes from around the lower boundary of the integral,
$\eta\approx\eta_0$, and so
corrections arising in the sub-horizon regime
are suppressed compared to those
imprinted beforehand in the super-horizon regime.

A similar expression for ${\cal E}'_2$
can be obtained as
\begin{eqnarray}
{\cal E}'_2=
-\frac{\ell^2}{8\az}\:
(\uz')^*\int d\omega
p^4\tildephiz e^{-i\omega\eta}\ln
\left( \frac{p^2}{k^2} \right)
+
\frac{2\ell^2}{\az^2}
\left[ \ln\left(\frac{k\ell}{2\az}\right)+\gamma \right]
\left[
\frac{3}{\eta^3} | \uz' |^2+
\left(-\frac{3}{\eta^4}+\frac{k^2}{\eta^2}\right)(\uz')^*\uz
\right] + \mbox{c.c.}.
\label{u-S_2}
\end{eqnarray}
The second term is local and suppressed for large 
$\eta$ for the same reason above. 
To examine the late time behavior of the first term,
we substitute the expression~(\ref{linear_combination})
without limitation $\eta>\eta_0$.
This approximation is justified because 
the kernel $\int d\omega e^{-i\omega(\eta-\eta')}
p^4 \ln (p^2/k^2)$ decays at least as fast as $1/(\eta-\eta')^3$ 
for a large time separation, as will be shown in Appendix. 
Under this approximation for $\phi(\eta)$, 
it is easy to calculate its Fourier transform
\footnote{
Depending on the choice of the integration path near $\eta=0$, 
the Fourier transform of $\phi(\eta)$ changes by a constant 
independent of $\omega$. 
%This is equivalent to adding a delta function 
%at $\eta=0$ to $\phi(\eta)$. 
Here we took the principal values. 
}
\begin{eqnarray}
\tildephiz (\omega) =
\frac{i \eta_0}{2a_0}\frac{1}{\sqrt{2k}}\times
\left\{
  \begin{array}{l}
       -(\alpha+\beta)~~~\mbox{for}~~~\omega\leq -k,\\
       (\alpha-\beta)~~~~~\mbox{for}~~~|\omega|\leq k,\\
       (\alpha+\beta)~~~~~\mbox{for}~~~\omega\geq k.
  \end{array}
\right.  
\end{eqnarray}
The contour of the $\omega$-integration in 
the first term of Eq.~(\ref{u-S_2}) is such shown 
in Fig. \ref{fig:cont1.eps},
and hence it can be separated into three parts as
\begin{eqnarray}
\mbox{The first term in (\ref{u-S_2})}=
-(\uz')^* \frac{k^2}{16} \frac{k^2\ell^2}{a_0\az}k\eta_0\left(
{\cal I}^{-k-i\infty}_{\to -k}+{\cal I}^{-k}_{\to k}+{\cal I}^{k}_{\to k-i\infty}
\right),
\label{int_res}
\end{eqnarray}
where
\begin{eqnarray}
&&{\cal I}^{k}_{\to k-i\infty}=({\cal I}^{-k-i\infty}_{\to -k})^*
=(\alpha+\beta)\frac{e^{-ik\eta}}{\sqrt{2k}}
\int^{\infty}_{0}dy(y^2+2iy)^2 e^{-k\eta\cdot y}
  \left[\ln(2y -iy^2)-{\pi i\over 2}\right],
\\
&&{\cal I}^{-k}_{\to k}=(\alpha-\beta)\frac{i}{\sqrt{2k}}
\int^{1}_{-1}dy(1-y^2)^2e^{-ik\eta \cdot y}\ln(1-y^2).
\end{eqnarray}
We can see that
all ${\cal I}$s are suppressed for large values of $k\eta$, 
since their integrand becomes a product of 
a sooth function and a rapidly oscillating function. 
The dominant contribution 
in the integration of ${\cal E'}$
therefore comes from around the lower boundary of the integral.

Based on the above analysis,
we conclude that all the dominant corrections arise in the 
super-horizon regime,
and thus it is sufficient to evaluate the corrections
in the long wavelength approximation.

%-----------------------------------------%
\section{Summary and Discussion}
%-----------------------------------------%
\label{sec:conclusion}

We have investigated leading order corrections
to the tensor perturbations in the RS II braneworld cosmology
by using the perturbative expansion scheme of Ref.~\cite{Tanaka:2004ig}.
We have studied a model composed of slow-roll inflation on the brane,
followed by a radiation dominant era. 
The unperturbed five dimensional bulk is AdS space 
with curvature radius $\ell$.
In our expansion scheme the asymptotic boundary 
conditions in the bulk are imposed by choosing 
outgoing solutions of bulk perturbations, whose general 
expression is known in the Poincar\'{e} coordinate system. 
Hence, the issue of bulk boundary conditions is handled without 
introducing an artificial regulator brane. 
This is one of the notable advantages of the present scheme.

We set the initial condition 
when the wave length $a k^{-1}$ is already longer 
than the Hubble radius during slow-roll inflation. 
We will not lose much by neglecting the modes which 
are already inside the Hubble scale at 
the time of transition to the radiation era, since 
they are not cosmologically so interesting. 
We compared the resulting amplitudes of fluctuations after 
the horizon re-entry
with and without the effect of an extra dimension.  
To do so, we normalized the amplitude 
so that the late time amplitude of 
fluctuations becomes identical in two cases 
if the slow-roll inflation lasts forever. 
As the reference without the effect of an extra dimension, 
we took a growing mode solution.    
The corrections due to gravity propagation through the fifth dimension 
dominantly come from the contribution around the transition time. 
Therefore they can be estimated by using the long wavelength 
approximation with a sufficient accuracy. 
Since the correction due to the modified cosmic expansion 
comes from relatively late epoch, it is necessary 
to take into account the oscillatory behavior of the 
solution after the re-entry to the Hubble horizon. 

Combining the results obtained in Eqs.~(\ref{corr_from_local}), 
(\ref{F2fin}) and (\ref{F0fin}),  
we find that the amplitude of a fluctuation with comoving 
wave number $k$ is 
modified by a factor $e^{\Re[F]}$ due to the effect of 
an extra dimension
with 
%and $\lim_{\eta\to\infty}\Re[F]$ is given by 
\begin{eqnarray}
\lim_{\eta\to \infty}\Re[F(\eta)]=
\frac{k^2\ell^2}{a_0^2}\left\{
\frac{13}{150}-\frac{4}{15}\left[\gamma+\ln\left(
\frac{k}{a_0H_0}
\right)\right]
+\frac{2}{5}\ln\left(\frac{\ell H_0}{2}\right)
\right\}, 
\end{eqnarray}
where $a_0$ and $H_0$ are the scale factor and the Hubble 
parameter at the transition time, and $\gamma$ is the Euler's constant. 
Leading corrections are proportional to $k^2\ell^2/a_0^2$ or 
$k^2\ell^2/a_0^2\log k$, as is expected from the dimensional 
analysis.  
However, our calculation here determined 
the precise numerical factors analytically.

Throughout this paper, we have dropped the contribution from the decaying mode 
for simplicity. 
Since the decaying mode during the initial slow-roll inflation 
phase is usually suppressed when the wavelength 
is longer than the Hubble scale, 
it does not give a significant effect. 
However, in the context of braneworld, we have not yet understood 
how to fix the initial conditions for fluctuations. 
Therefore, in principle, there is a possibility that 
the contamination of the decaying mode can be extremely large,
although it seems quite unlikely. 
%already of $O(\ell^2)$, 
%its contribution to the correction from the $E_{\mu\nu}$ term 
%due to the right hand side of Eq.~(\ref{iteration}) 
%becomes higher order. 
In such a case the contribution from the decaying component cannot 
be neglected. 

It is quite easy to 
estimate the leading order correction due to the decaying mode, 
since it does not pick up the 
effect of the presence of an extra dimension. 
Here the effect is not due to the non-trivial evolution 
of a solution but totally due to unconventional initial condition. 
Adding a decaying mode will modify $v$ given in Eq.~(\ref{dsmodefunc}) to
\begin{eqnarray*}
v(\eta)\simeq {A_ka_0\eta_0\over 2\eta_0 -\eta}
  \left[{1}
   +{k^2(2\eta_0-\eta)^2\over 2} 
   +C_d {k^3(2\eta_0-\eta)^3}\right], 
\end{eqnarray*}
where $C_d$ is a complex number which parametrizes the 
amplitude of the decaying component. 
Repeating a similar calculation leading to Eq.~(\ref{F0fin}), we find 
that the relative change in the amplitude due to  the decaying mode 
is given by 
\begin{eqnarray}
 \Re[C_d] {\ell^2 H_0^2\over 2}(k\eta_0)^3.
\end{eqnarray}
Thus the correction due to the decaying component is more 
suppressed in the sense of the power of $k$, but 
it might be more important than the other corrections 
for $k\eta_0 > \Re[C_d]^{-1}$ if $\Re[C_d]$ is extremely large.  
%Since $C_d$Here dimensional variables 
%are $H_0$ and $\ell$, If it is the case, 
%the expression for $C_d$ should 
%have $\ell$ in the denominator
%$C_d=O(\epsilon/H_0^2\ell^2)$

Lastly we would like to mention the limitation of the present 
work and some future issues.
The perturbative expansion scheme that we have adopted in this paper
is valid for scales larger than the 
AdS curvature length ($k\ell /a\ll 1$)
at low energies ($\ell H \ll 1$);
besides our study is
restricted to initially super-horizon perturbations.
We gave an initial condition for a solution by hand, 
but of course such an approach is not satisfactory. 
To determine the initial condition, 
we need to investigate the evolution of perturbations 
at the initial phase where the wavelength of the mode is 
too short and reduction to an effective four dimensional
problem is not possible. 
Furthermore, 
to seek for interesting effects that might arise at 
high energies ($\ell H \gtrsim 1$),
we have to develop a new formalism, which is a next challenge
and will be hopefully reported in our future publication.

\acknowledgments
We are grateful to Shinji Mukohyama and Roy Maartens 
for useful discussions. This work is supported in part 
by Monbukagakusho Grant-in-Aid Nos.
16740141 and 14047212, Inamori Foundation, 
and 21COE program at Kyoto university
``Center for Diversity and Universality in Physics''.

%This work is partly supported by 
%Monbukagakusho Grant-in-Aid Nos. 12740154 and 
%and by Inamori foundation. 

\appendix
\section{Details of calculations}

In this appendix, we will show the following formula:
\begin{eqnarray}
\int d\omega e^{-i\omega T}\ln\left( \frac{p^2}{k^2}\right)
=-4\pi \left\{ \theta (T) \frac{\cos(kT)-1}{T}
+\partial_T\left[ \theta (T) \ln (kT) \right]+ \gamma\delta (T) \right\},
\label{app_formula}
\end{eqnarray}
where $p^2= k^2 - \omega^2$ and $\gamma$ is Euler's constant.
The branch cuts are all running on the lower half complex plane of $\omega$
(Fig.\ref{fig:cont1.eps}), which ensures the retarded boundary conditions.

\begin{figure}[htb]
  \begin{center}
    \includegraphics[keepaspectratio=true,height=50mm]{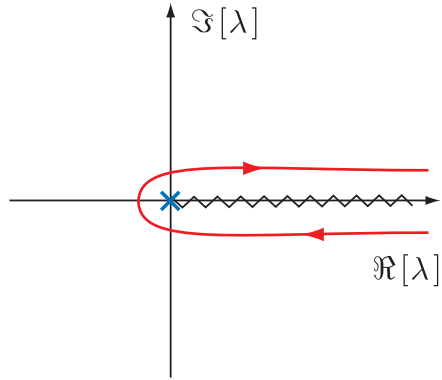}
  \end{center}
  \caption{Branch cut and the contour of the integration in Eq.~(\ref{app_lambda}).}
  \label{fig:cont2.eps}
\end{figure}

We divide the above integration into two parts:
\begin{eqnarray}
\int d\omega e^{-i\omega T}\ln \left(\frac{p^2}{k^2}\right)
=\int d\omega e^{-i\omega T}\ln\left(\frac{-p^2}{\omega^2}\right)
+\int d\omega e^{-i\omega T}\ln\left(\frac{-\omega^2}{k^2}\right).
\label{app_two_parts}
\end{eqnarray}
The first term of the right hand side can be rewritten as
\begin{eqnarray*}
\int d\omega e^{-i\omega T}\ln\left(\frac{-p^2}{\omega^2}\right)
=\int  d\omega e^{-i\omega T}
\left[\ln (\omega-k)+\ln (\omega+k)-2 \ln \omega \right].
\end{eqnarray*}
The integration can be performed as
\begin{eqnarray*}
\int d\omega e^{-i\omega T} \ln (\omega-k)
&=& \theta(T)\left[
\int^{0}_{\infty}(-i)dr e^{-iT(k-ir)}\ln(r e^{3\pi i/2})
+\int^{\infty}_{0}(-i)dr e^{-iT(k-ir)}\ln(r e^{-\pi i/2})
\right]
\\
&=&-2\pi \theta(T)\frac{e^{-ikT}}{T},
\end{eqnarray*}
and the other two integrals are obtained by setting $k\to -k$ and $k\to 0$.
Thus we have
\begin{eqnarray*}
\int d\omega e^{-i\omega T}\ln\left(\frac{-p^2}{\omega^2}\right)
=-4\pi\theta(T)\frac{\cos(kT)-1}{T}.
\end{eqnarray*}
This expression is regular at $T=0$.
To evaluate the second term in the right hand side of Eq.~(\ref{app_two_parts}),
we need a trick
because the integrand suffers from a slow fall-off at large values of $|\omega|$.
It should be interpreted as
\begin{eqnarray}
2 i \partial_T \left[\int d\omega
\frac{e^{-i\omega T}}{\omega}\ln\left(\frac{-i \omega}{k}\right)\right]
=2 i \partial_T \left\{
\theta(T)\left[
\int d\lambda\frac{e^{-\lambda}}{\lambda}\left( \ln(- \lambda) - \ln (kT)\right)
\right]\right\},
\label{app_lambda}
\end{eqnarray}
where we put $\lambda = i\omega T$.
Now the branch cut and the contour of the integration on $\lambda$ plane
are such shown in Fig~\ref{fig:cont2.eps}.
Thanks to this trick, the expression for the second term
becomes a well-defined one, though the result does not change
as long as convergence is guaranteed.
Then, performing the integration as
\begin{eqnarray*}
\int d\lambda\frac{e^{-\lambda}}{\lambda}\left( \ln (-\lambda) - \ln (kT)\right)
&=&\lim_{\nu\to 0}\partial_{\nu}
\int d\lambda e^{-\lambda}(-\lambda)^{\nu -1}+2\pi i \ln(kT)
\\
&=&\lim_{\nu\to 0}\partial_{\nu}\left[ (e^{-\pi i \nu} -e^{\pi i \nu})
\Gamma (\nu) \right] +2\pi i \ln(kT)
\\
&=&2\pi i \left[ \ln (kT) + \gamma \right],
\end{eqnarray*}
we finally obtain the formula~(\ref{app_formula}).

For $T>0$ the formula~(\ref{app_formula}) reduces to 
$
\int d\omega e^{-i\omega T}\ln\left(p^2/k^2\right)
=-4\pi \cos(kT)/ T 
$
and hence the kernel that appears in ${\cal S}_2$ becomes 
\begin{equation}
\int d\omega e^{-i\omega T} p^4 \ln\left( \frac{p^2}{k^2}\right)
=32\pi \left[
   \left({k^2\over T^3}-{3\over T^5}\right)\cos( kT)
         -{3k\over T^4} \sin(kT) \right],
\qquad \mbox{for}~~T>0. 
\label{p4ln}
\end{equation}
From this expression we find that the non-local source ${\cal S}_2$ 
does not keep the past history for a long time.

%---------   References   ---------%

%---------   References   ---------%
%-------------------------------------------------------------%

\begin{thebibliography}{99}

%\cite{Rubakov:2001kp}
\bibitem{Rubakov:2001kp}
V.~A.~Rubakov,
%``Large and infinite extra dimensions: An introduction,''
Phys.\ Usp.\  {\bf 44}, 871 (2001)
[Usp.\ Fiz.\ Nauk {\bf 171}, 913 (2001)]
[arXiv:hep-ph/0104152].
%%CITATION = HEP-PH 0104152;%%

%\cite{Langlois:2002bb}
\bibitem{Langlois:2002bb}
D.~Langlois,
%``Brane cosmology: An introduction,''
Prog.\ Theor.\ Phys.\ Suppl.\  {\bf 148}, 181 (2003)
[arXiv:hep-th/0209261].
%%CITATION = HEP-TH 0209261;%%

%\cite{Maartens:2003tw}
\bibitem{Maartens:2003tw}
R.~Maartens,
``Brane-world gravity,''
arXiv:gr-qc/0312059.
%%CITATION = GR-QC 0312059;%%

%\cite{Randall:1999vf}
\bibitem{Randall:1999vf}
L.~Randall and R.~Sundrum,
%``An alternative to compactification,''
Phys.\ Rev.\ Lett.\  {\bf 83}, 4690 (1999)
[arXiv:hep-th/9906064].

%\cite{Garriga:1999yh}
\bibitem{Garriga:1999yh}
J.~Garriga and T.~Tanaka,
%``Gravity in the brane-world,''
Phys.\ Rev.\ Lett.\  {\bf 84}, 2778 (2000)
[arXiv:hep-th/9911055].
%%CITATION = HEP-TH 9911055;%%

\bibitem{Gidding}
S.~B.~Giddings, E.~Katz and L.~Randall,
%``Linearized gravity in brane backgrounds,''
JHEP {\bf 0003}, 023 (2000)
[arXiv:hep-th/0002091].
%%CITATION = HEP-TH 0002091;%%


%---Refs for Cosmological Pertrubations ---%

%\cite{Maartens:1999hf}
\bibitem{Maartens:1999hf}
R.~Maartens, D.~Wands, B.~A.~Bassett and I.~Heard,
%``Chaotic inflation on the brane,''
Phys.\ Rev.\ D {\bf 62}, 041301 (2000)
[arXiv:hep-ph/9912464].
%%CITATION = HEP-PH 9912464;%%


%\cite{Mukohyama:2000ui}
\bibitem{Mukohyama:2000ui}
S.~Mukohyama,
 %``Gauge-invariant gravitational perturbations of maximally symmetric
%spacetimes,''
Phys.\ Rev.\ D {\bf 62}, 084015 (2000)
[arXiv:hep-th/0004067].
%%CITATION = HEP-TH 0004067;%%

%\cite{Kodama:2000fa}
\bibitem{Kodama:2000fa}
H.~Kodama, A.~Ishibashi and O.~Seto,
%``Brane world cosmology: Gauge-invariant formalism for perturbation,''
Phys.\ Rev.\ D {\bf 62}, 064022 (2000)
[arXiv:hep-th/0004160].
%%CITATION = HEP-TH 0004160;%%

%\cite{Langlois:2000ia}
\bibitem{Langlois:2000ia}
D.~Langlois,
%``Brane cosmological perturbations,''
Phys.\ Rev.\ D {\bf 62}, 126012 (2000)
[arXiv:hep-th/0005025].
%%CITATION = HEP-TH 0005025;%%

%\cite{Langlois:2000ph}
\bibitem{Langlois:2000ph}
D.~Langlois,
%``Evolution of cosmological perturbations in a brane-universe,''
Phys.\ Rev.\ Lett.\  {\bf 86}, 2212 (2001)
[arXiv:hep-th/0010063].
%%CITATION = HEP-TH 0010063;%%

%\cite{Koyama:2000cc}
\bibitem{Koyama:2000cc}
K.~Koyama and J.~Soda,
%``Evolution of cosmological perturbations in the brane world,''
Phys.\ Rev.\ D {\bf 62}, 123502 (2000)
[arXiv:hep-th/0005239].
%%CITATION = HEP-TH 0005239;%%

%\cite{Koyama:2001ct}
\bibitem{Koyama:2001ct}
K.~Koyama and J.~Soda,
%``Bulk gravitational field and cosmological perturbations on the brane,''
Phys.\ Rev.\ D {\bf 65}, 023514 (2002)
[arXiv:hep-th/0108003].
%%CITATION = HEP-TH 0108003;%%

%\cite{Koyama:2003be}
\bibitem{Koyama:2003be}
K.~Koyama,
%``CMB anisotropies in brane worlds,''
Phys.\ Rev.\ Lett.\  {\bf 91}, 221301 (2003)
[arXiv:astro-ph/0303108].
%%CITATION = ASTRO-PH 0303108;%%

%--- GWs ---%

%\cite{Langlois:2000ns}
\bibitem{Langlois:2000ns}
D.~Langlois, R.~Maartens and D.~Wands,
%``Gravitational waves from inflation on the brane,''
Phys.\ Lett.\ B {\bf 489}, 259 (2000)
[arXiv:hep-th/0006007].
%%CITATION = HEP-TH 0006007;%%

%\cite{Gorbunov:2001ge}
\bibitem{Gorbunov:2001ge}
D.~S.~Gorbunov, V.~A.~Rubakov and S.~M.~Sibiryakov,
%``Gravity waves from inflating brane or mirrors moving in adS(5),''
JHEP {\bf 0110}, 015 (2001)
[arXiv:hep-th/0108017].
%%CITATION = HEP-TH 0108017;%%

%\cite{Kobayashi:2003cn}
\bibitem{Kobayashi:2003cn}
T.~Kobayashi, H.~Kudoh and T.~Tanaka,
%``Primordial gravitational waves in inflationary braneworld,''
Phys.\ Rev.\ D {\bf 68}, 044025 (2003)
[arXiv:gr-qc/0305006].
%%CITATION = GR-QC 0305006;%%

%\cite{Hiramatsu:2003iz}
\bibitem{Hiramatsu:2003iz}
T.~Hiramatsu, K.~Koyama and A.~Taruya,
 %``Evolution of gravitational waves from inflationary brane-world : Numerical
%study of high-energy effects,''
Phys.\ Lett.\ B {\bf 578}, 269 (2004)
[arXiv:hep-th/0308072].
%%CITATION = HEP-TH 0308072;%%

%\cite{Easther:2003re}
\bibitem{Easther:2003re}
R.~Easther, D.~Langlois, R.~Maartens and D.~Wands,
%``Evolution of gravitational waves in Randall-Sundrum cosmology,''
JCAP {\bf 0310}, 014 (2003)
[arXiv:hep-th/0308078].
%%CITATION = HEP-TH 0308078;%%

%\cite{Battye:2003ks}
\bibitem{Battye:2003ks}
R.~A.~Battye, C.~Van de Bruck and A.~Mennim,
 %``Cosmological tensor perturbations in the Randall-Sundrum model: Evolution in
%the near-brane limit,''
Phys.\ Rev.\ D {\bf 69}, 064040 (2004)
[arXiv:hep-th/0308134].
%%CITATION = HEP-TH 0308134;%%


%\cite{Ichiki:2003hf}
\bibitem{Ichiki:2003hf}
K.~Ichiki and K.~Nakamura,
``Causal structure and gravitational waves in brane world cosmology,''
arXiv:hep-th/0310282.
%%CITATION = HEP-TH 0310282;%%

%\cite{Ichiki:2004sx}
\bibitem{Ichiki:2004sx}
K.~Ichiki and K.~Nakamura,
``Stochastic gravitational wave background in brane world cosmology,''
arXiv:astro-ph/0406606.
%%CITATION = ASTRO-PH 0406606;%%

%\cite{Koyama:2004cf}
\bibitem{Koyama:2004cf}
K.~Koyama,
``Late time behavior of cosmological perturbations in a single brane model,''
arXiv:astro-ph/0407263.
%%CITATION = ASTRO-PH 0407263;%%

%--- GWs ---%

%\cite{Koyama:2003yz}
\bibitem{Koyama:2003yz}
K.~Koyama and K.~Takahashi,
%``Primordial fluctuations in bulk inflaton model,''
Phys.\ Rev.\ D {\bf 67}, 103503 (2003)
[arXiv:hep-th/0301165].
%%CITATION = HEP-TH 0301165;%%

%\cite{Koyama:2003sb}
\bibitem{Koyama:2003sb}
K.~Koyama and K.~Takahashi,
 %``Exactly solvable model for cosmological perturbations in dilatonic brane
%worlds,''
Phys.\ Rev.\ D {\bf 68}, 103512 (2003)
[arXiv:hep-th/0307073].
%%CITATION = HEP-TH 0307073;%%

%\cite{Kobayashi:2003cb}
\bibitem{Kobayashi:2003cb}
T.~Kobayashi and T.~Tanaka,
%``Bulk inflaton shadows of vacuum gravity,''
Phys.\ Rev.\ D {\bf 69}, 064037 (2004)
[arXiv:hep-th/0311197].
%%CITATION = HEP-TH 0311197;%%

%\cite{Binetruy:2004dw}
\bibitem{Binetruy:2004dw}
P.~Binetruy, M.~Bucher and C.~Carvalho,
``Models for the brane-bulk interaction: Toward understanding braneworld
%cosmological perturbations,''
arXiv:hep-th/0403154.
%%CITATION = HEP-TH 0403154;%%


%---Refs for Cosmological Pertrubations ---%


%\cite{Tanaka:2004ig}
\bibitem{Tanaka:2004ig}
T.~Tanaka,
``AdS/CFT correspondence in a Friedmann-Lemaitre-Robertson-Walker brane,''
arXiv:gr-qc/0402068.

%\cite{Binetruy:1999ut}
\bibitem{Binetruy:1999ut}
P.~Binetruy, C.~Deffayet and D.~Langlois,
%``Non-conventional cosmology from a brane-universe,''
Nucl.\ Phys.\ B {\bf 565}, 269 (2000)
[arXiv:hep-th/9905012].
%%CITATION = HEP-TH 9905012;%%

%\cite{Binetruy:1999hy}
\bibitem{Binetruy:1999hy}
P.~Binetruy, C.~Deffayet, U.~Ellwanger and D.~Langlois,
%``Brane cosmological evolution in a bulk with cosmological constant,''
Phys.\ Lett.\ B {\bf 477}, 285 (2000)
[arXiv:hep-th/9910219].
%%CITATION = HEP-TH 9910219;%%

%\cite{Mukohyama:1999qx}
\bibitem{Mukohyama:1999qx}
S.~Mukohyama,
%``Brane-world solutions, standard cosmology, and dark radiation,''
Phys.\ Lett.\ B {\bf 473}, 241 (2000)
[arXiv:hep-th/9911165].
%%CITATION = HEP-TH 9911165;%%

%\cite{Ida:1999ui}
\bibitem{Ida:1999ui}
D.~Ida,
%``Brane-world cosmology,''
JHEP {\bf 0009}, 014 (2000)
[arXiv:gr-qc/9912002].
%%CITATION = GR-QC 9912002;%%

%\cite{Kraus:1999it}
\bibitem{Kraus:1999it}
P.~Kraus,
%``Dynamics of anti-de Sitter domain walls,''
JHEP {\bf 9912}, 011 (1999)
[arXiv:hep-th/9910149].
%%CITATION = HEP-TH 9910149;%%


%\cite{Shiromizu:1999wj}
\bibitem{Shiromizu:1999wj}
T.~Shiromizu, K.~i.~Maeda and M.~Sasaki,
%``The Einstein equations on the 3-brane world,''
Phys.\ Rev.\ D {\bf 62}, 024012 (2000)
[arXiv:gr-qc/9910076].
%%CITATION = GR-QC 9910076;%%

%%<----- Gradient expansion
%\cite{Kanno:2002ia}
\bibitem{KannoSoda}
S.~Kanno and J.~Soda,
%``Brane world effective action at low energies and AdS/CFT,''
Phys.\ Rev.\ D {\bf 66}, 043526 (2002)
[arXiv:hep-th/0205188].
%%CITATION = HEP-TH 0205188;%%

%\cite{Kanno:2002ia}
\bibitem{Kanno:2002ia}
S.~Kanno and J.~Soda,
%``Radion and holographic brane gravity,''
Phys.\ Rev.\ D {\bf 66}, 083506 (2002)
[arXiv:hep-th/0207029].
%%CITATION = HEP-TH 0207029;%%

%\cite{Koyama:2002nw}
\bibitem{Koyama:2002nw}
K.~Koyama,
%``Radion and large scale anisotropy on the brane,''
Phys.\ Rev.\ D {\bf 66}, 084003 (2002)
[arXiv:gr-qc/0204047].
%%CITATION = GR-QC 0204047;%%

%% Gradient expansion ----->

\end{thebibliography}
\end{document}